# Three-dimensional Character of the Magnetization Dynamics in Magnetic Vortex Structures - Hybridization of Flexure Gyromodes with Spin Waves

Matthias Noske,[1*] Hermann Stoll,[1] Manfred Fähnle,[1] Ajay Gangwar,[2] Georg Woltersdorf,[3] Andrei Slavin,[4] Markus Weigand,[1] Georg Dieterle,[1] Johannes Förster,[1] Christian H. Back[2] and Gisela Schütz[1]

[1]*Max Planck Institute for Intelligent Systems, Heisenbergstr. 3, 70569 Stuttgart, Germany*
[2]*University of Regensburg, Department of Physics, Universitätsstraße 31, 93053 Regensburg, Germany*
[3]*University of Halle, Department of Physics, Von-Danckelmann-Platz 3, 06120 Halle, Germany*
[4]*Oakland University, Department of Physics, Rochester, Michigan 48309, USA*

Three-dimensional linear spin-wave eigenmodes of a Permalloy disk having finite thickness are studied by micromagnetic simulations based on the Landau-Lifshitz-Gilbert equation. The eigenmodes found in the simulations are interpreted as linear superpositions (hybridizations) of 'approximate' three-dimensional eigenmodes, which are the fundamental gyromode $G_0$, the spin-wave modes and the higher-order gyromodes $G_N$ (flexure modes), the thickness dependence of which is represented by perpendicular standing spin waves. This hybridization leads to new and surprising dependencies of the mode frequencies on the disk thickness. The three-dimensional character of the eigenmodes is essential to explain the recent experimental results on vortex-core reversal observed in relatively thick Permalloy disks.

PACs numbers: 75.75.Jn, 75.30.Ds, 75.78.Cd

---

[*]Corresponding author: noske@is.mpg.de

In a cylindrical nanodisk with thickness $h$ of a few tens of nm and diameter $2R$ of typically several hundred nm the magnetic ground state is a vortex. There the magnetization curls in the plane of the disk with a clockwise (CW) circulation ($c = -1$) or counterclockwise (CCW) circulation ($c = +1$). At the center of the disk in an area with a typical diameter of 10 to 20 nm the magnetization turns out of the plane [1] forming the vortex core, which points either up or down, corresponding to the two polarities $p = +1$ and $p = -1$.

There are various eigenmodes of the magnetization dynamics in such systems containing a vortex. In an analytical approach "approximate" eigenmodes are obtained from a simplified version of the linearized Landau-Lifshitz-Gilbert (LLG) equation [2,3]. The simplification is that the non-diagonal part of the dipole-dipole interaction appearing in the effective field of the LLG equation is neglected [4].

In the two-dimensional case of a thin vortex-state disk (i.e. for a thickness of only a few exchange lengths $l_{ex} = \sqrt{2A/(\mu_0 M_s^2)}$ (in SI units [5]) the "approximate" eigenmodes are the fundamental gyromode $G_0$ and the dipolar spin-wave modes [6-9]. In thicker disks the "approximate" eigenmodes can have a non-trivial variation of the dynamical magnetization along the disk thickness, and can include so-called "higher order gyromodes" $G_N$ (or "flexure" modes) [10-13].

The fundamental mode $G_0$ describes the circular motion of the vortex core typically in the range of 100 MHz to 1 GHz. The dipolar in-plane modes describe dipolar excitations of the planar part of the vortex and are denoted by a pair of integers (*n,m*). The radial mode number $n \geq 1$ is the number of radial nodes in the out-of-plane component of the dynamical magnetization. The azimuthal mode number m can have negative and positive values and |2m| is the number of the azimuthal nodes. The frequency $\omega_{p,n,m}$ depends on the mode numbers n and m and on the core polarity $p$, specifically on the product $pm$ and is typically in the multi-GHz frequency range. In the following we consider only the azimuthal in-plane modes with $|m| = 1$.

For disk thicknesses above a few exchange lengths all the "approximate" eigenmodes are three-dimensional. The lowest three-dimensional modes of a vortex-state magnetic disk have the in-plane magnetization distribution similar to the mode $G_0$, and are characterized by flexure oscillations of the vortex core line along the thickness of the disk (Fig.1). The mode number $N \geq 1$ indicates the number of nodes.

In the literature, so far, only the linear superposition (hybridization) of the $n = 1, m = \pm 1$ spin-wave modes and the fundamental mode $G_0$ in thin magnetic disks has been investigated by analytical calculations. This hybridization leads to the frequency splitting of the azimuthal dipolar spin waves [9].

In the present letter we investigate the numerically simulated "true" eigenmodes of thick magnetic disks. These eigenmodes can be interpreted as hybridizations between the "approximate" (or diagonal) three-dimensional eigenmodes due to the non-diagonal dipolar interaction between them. In order to remain in the linear regime, only small amplitude excitations are used to analyze the eigenmodes. Here we especially focus on dipolar spin waves where three-dimensionality has not been discussed so far. We found that besides a change of the mode profiles these modes can hybridize with the inherently three-dimensional flexure modes. This effect strongly changes the frequency behavior of the spin-wave modes, including a frequency gap in the crossing region of the modes. The simulations are performed for circular Permalloy disks (see supplementary information).

Fig. 2 shows the frequencies of the vortex eigenmodes as a function of the thickness $h$. The respective dominant character of the eigenmode is given by the color of the line: It is orange for $G_0$, red for $G_1$, blue for the $n = 1, m = -1$ azimuthal spin wave and green for the $n = 1, m = +1$ spin wave. For a vortex pointing up the senses of rotation of these modes are CCW (when looking from the top) for $G_0$, $G_1$ and for the $n = 1, m = 1$ mode and CW for the $n = 1, m = -1$ mode.

For the low-frequency gyromode $G_0$ the magnetization $\boldsymbol{M}(\boldsymbol{r}, t)$ is nearly independent of $z$, i.e., it is a uniform (even) mode. For the pure unhybridized $G_0$ mode the frequency increases approximately in a linear manner [7] with increasing $h$ (for fixed diameter $2R$), for larger values of $h$ the increase is slightly smaller. The gyromode hybridizes [6,9] rather strongly with azimuthal spin waves with $n = 1$, $m = \pm 1$ which also have an even $\boldsymbol{M}(\boldsymbol{r}, t)$. The hybridization leads to the frequency splitting discussed in the introduction. It also leads to a renormalized frequency of the gyromode [6] which is slightly smaller than the one of the unhybridized mode. Micromagnetic simulations (see Fig. 2, orange curve and Figure 3 of ref. [10]) show that in reality the frequency of the gyromode is smaller than predicted by the theory [6,10]. Possibly this difference arises from a repulsion between the gyromode $G_0$ and the flexure mode $G_1$, however so far no corresponding theory exists.

A hybridized mode with dominant character of the flexure mode $G_1$ is denoted by the red color in Fig. 2. Approximate analytical calculations [10] predict that for large thickness $h$ the frequency of the non-hybridized $G_1$ mode should scale as $h^{-2}$ which is similar to the dependence we found in our simulations. For small $h$ this mode is less relevant as it requires too much exchange energy. There are two sections of the mode with dominant $G_1$ character appearing on two separate curves I and II in Fig. 2. One section occurs for large $h$ down to about $h = 40$ nm (curve I, red segment). The second section with $G_1$ character is located on curve II (red segment) for $h < 40$ nm.

The $n = 1, m = +1$ azimuthal spin-wave mode is located on the other segments of curves I and II (Fig. 2, green segments). For thin disks, this mode is located on curve I and for large thicknesses it is located on curve II. This means that the dominant character of curve I changes from an azimuthal spin-wave mode ($n = 1, m = +1$) towards $G_1$ with increasing disk thickness (and vice versa for curve II). Altogether, at the region where the non-hybridized $G_1$ mode and the non-hybridized $n = 1, m = +1$ mode would cross, the hybridization causes a frequency gap between curves I and II.

Another finding is that the type of hybridization for the $n = 1, m = +1$ mode shows a dependence on disk thickness which has not been discussed in literature so far. For thin disks, hybridization occurs with the fundamental gyromode $G_0$ as can be seen from the vortex core gyration radius which is nearly independent of $z$ (Fig. 2f) and the dynamic mode profile (Fig. 3f). This is in agreement with ref. [9]. However, at medium thicknesses $h$ this spin-wave mode hybridizes first with $G_1$, as can be seen by the minimum in the vortex core gyration radius (Fig. 2b) and then dominantly with $G_2$ (Fig. 2c: two minima in the core gyration radius). This means there is a change of the dominant hybridization from $G_0$ to $G_1$ to $G_2$ with increasing thickness.

The $n = 1, m = -1$ azimuthal spin wave (Fig. 2, blue curve) has lower frequencies [9] than the mode with dominant $n = 1, m = +1$ character. In contrast to the $n = 1, m = +1$ mode, for the $n = 1, m = -1$ mode the vortex core gyration radius $r(z)$ is nearly $z$-independent for every disk thickness (Figs. 2d-f). This means the $n = 1, m = -1$ mode hybridizes dominantly with $G_0$ independently of disk thickness. The mode hybridizes only weakly with $G_1$ because $G_1$ is an odd mode and because the senses of rotation are different (CW for the $n = 1, m = -1$ mode and CCW for $G_1$), while they are the same for the CCW rotating $n = 1, m = +1$ mode. The $G_2$ mode is also even, but its sense of rotation is CCW, and therefore the hybridization between the $n = 1, m = -1$ spin wave and $G_2$ is also weak.

At a large thickness of about $h = 100$ nm the higher order spin-wave modes $n = (2,3,4,...), m = +1$ with the same sense of rotation as the vortex gyrotropic modes all hybridize with $G_2$, whereas for the $m = -1$ modes with an opposite sense of rotation only hybridization with $G_0$ is found.

We have also performed simulations for rings where the central part of the vortex structure is removed so that there is no vortex core and thus no gyrotropic eigenmodes $G_N$. As a result, the spin-wave eigenmodes are degenerate for $m = \pm|1|$. The eigenfrequency of the $n = 1, m = \pm 1$ mode in the ring is plotted in Fig. 2 (black line). It is always located between the eigenfrequencies of the $n = 1, m = +1$ and $n = 1, m = -1$ modes of the system with a vortex core. This means that the non-monotonic behavior of the frequency of the dipolar azimuthal modes in a magnetic disk, either with or without the vortex core, is not related to their hybridization with either of the gyrotropic modes.

An analysis of the dynamic mode profiles reveals a z-dependence outside the core region,

showing that also the non-hybridized spin waves have three-dimensional character (Fig. 3 b, c, e, f perspective views). The spin oscillation amplitude has a maximum at $z = 1/2h$. This is similar to the profile of exchange-dominated perpendicular standing spin waves (PSSWs). For a general disk thickness the z-dependent mode profile of the spin wave can be represented as a product of the mode profile of a two-dimensional spin wave (which has been calculated analytically for systems with extremely thin disks) with a z-dependent function $f(z)$. Thereby $f(z)$ can be represented as a Fourier series in cosines of z, which is constructed in such a way, that $f(z)$ has the correct number of nodes in the z-direction. The individual Fourier components, thereby, have the form of PSSWs.

The profile of the mode $G_0$, then, can be formally represented as a product of the analytically known profile calculated for extremely thin films with the uniform thickness profile (described by a cosine with argument 0, i.e. a constant).

So far, no analytical calculations of the PSSWs in magnetic vortex structures exist, although, the eigenfrequencies of the PSSWs have been calculated for infinite in-plane magnetic films [14]. There, the PSSW eigenfrequency is inversely proportional to the square of the film thickness ($\sim 1/h^2$).

Due to the exchange-dominated character of the PSSWs, a similar frequency behavior is expected from the non-trivial three-dimensional eigenmodes in vortex-state magnetic structures, which would explain the decrease of the "true" numerically calculated eigenfrequencies observed with increasing disk thickness $h$.

In contrast, the hybridization of the dipolar azimuthal modes ($n = 1, m = \pm 1$) with the lowest gyrotropic mode $G_0$ leads, as discussed above, to the splitting of their frequencies [9], and the hybridization of these modes with the gyrotropic flexure mode $G_1$ leads to the pronounced repulsion between $G_1$ and the $m = +1$ mode shown in Fig. 2, and, most likely, to the decrease in frequency splitting at larger disk thicknesses, since that splitting is originally caused by the hybridization with the $G_0$ mode [9].

Recently experiments on vortex core reversal by pulsed excitation of spin waves have been performed [15]. The experimental results could not be reproduced by two-dimensional micromagnetic simulations but only by three-dimensional simulations. It was assumed that the reason for this are z-dependent vortex core trajectories found in the three-dimensional simulations. In the present letter we have shown that the three-dimensionality of the vortex core dynamics is strongly influenced by the hybridization of the azimuthal spin waves and higher order gyromodes and therefore suggest that this hybridization has a significant influence on the switching experiments shown in [15].

To conclude, we have performed micromagnetic simulations of the linear dynamical behavior of a vortex-state cylindrical Permalloy disk having a finite thickness based on the Landau-Lifshitz-Gilbert equation. The numerically calculated spin-wave eigenmodes of the disk are interpreted in terms of hybridized "approximate" three-dimensional eigenmodes of a simplified version of the equation of motion, where the non-diagonal part of the dipole-dipole interaction has been neglected. By this we have investigated the hybridization between spin-wave modes and the fundamental gyrotropic mode $G_0$ and higher-order gyrotropic flexure modes $G_N$, $N \geq 1$ which have an essential three-dimensional character. New aspects of the hybridizations have been discovered. The hybridization with the $G_0$ mode leads to the well-known frequency splitting of the $n = 1, m = \pm 1$ azimuthal spin-wave modes, whereas the hybridization with higher order modes modifies the mode frequencies of the disk as soon as there are significant contributions of the flexure modes and introduces a frequency gap between the hybridized modes in the region where the unhybridized $G_1$ and $n = 1, m = +1$ azimuthal spin-wave modes would cross each other. For disks with a thickness exceeding a few exchange lengths of the disk material all the "approximate" eigenmodes are three-dimensional.

The authors are indebted to Riccardo Hertel for valuable discussions.

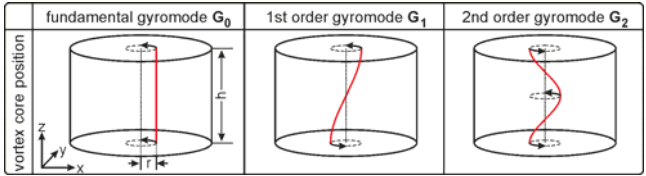

**Figure 1**: Schematic representation of the gyrotropic modes $G_0$, $G_1$ and $G_2$ (from left to right). The blue dots indicate nodes of the vortex core gyration radius.

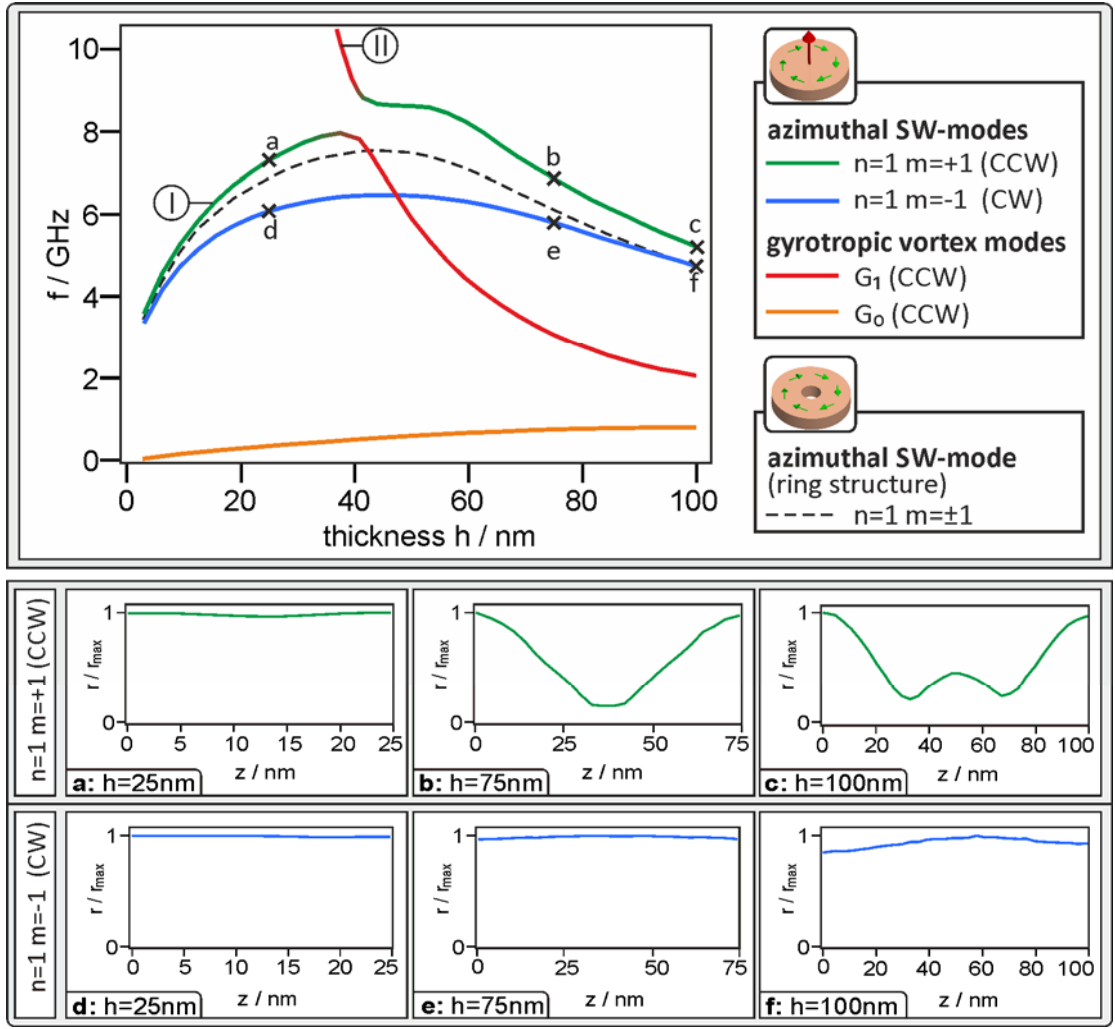

**Figure 2**: Eigenfrequencies $f_N$ of the vortex gyrotropic eigenmodes $G_0$ and $G_1$ and the azimuthal spin-wave modes $n=1, m=\pm 1$ as function of the disk thickness $h$ obtained from micromagnetic simulations by Fourier analysis (see text). The gyration radius of the vortex core (a-f) shows that the azimuthal SW-modes do not only hybridize with $G_0$(a, d-f), but also with $G_1$(b) and $G_2$(c). This is confirmed by the dynamic mode profiles corresponding to the points a-f shown in Fig. 3.

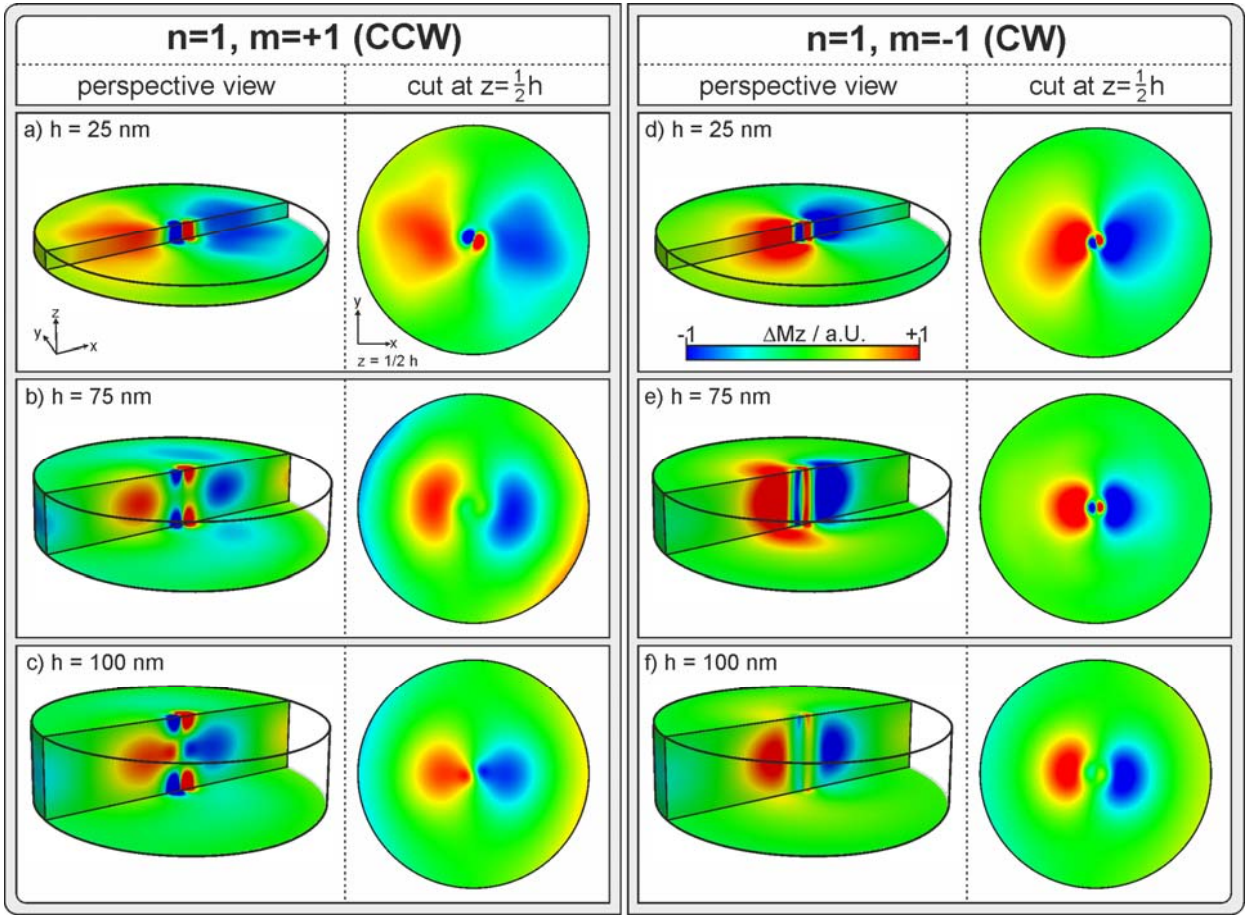

**Figure 3**: Analysis of the characters of the modes for the points a-f marked in Figure 2. All cuts at $z = 1/2h$ show the bipolar structure typical for a $n = 1, m = \pm 1$ azimuthal spin-wave mode. The perspective representations show that at the core region there is no minimum in the precession amplitude in a and d-f, one minimum in b and two minima in c, indicating a hybridization with $G_0$, $G_1$ and $G_2$, respectively.

# References

[1] A. Wachowiak, J. Wiebe, M. Bode, O. Pietzsch, M. Morgenstern, and R. Wiesendanger, Science **298**, 577 (2002).
[2] L. D. Landau and E. M. Lifshitz, Phys. Z. Sowjetunion **8**, 153 (1935).
[3] T. L. Gilbert, IEEE Trans. Magn. **40**, 3443 (2004).
[4] B. A. Kalinikos and A. N. Slavin, J. Phys. C **19**, 7013 (1986).
[5] G. S. Abo, Y.-K. Hong, J. Park, J. Lee, W. Lee, and B.-C. Choi, IEEE Trans. Magn. **49**, 4937 (2013).
[6] K. Y. Guslienko, G. R. Aranda, and J. M. Gonzalez, Phys. Rev. B **81**, 014414 (2010).
[7] K. Y. Guslienko, B. A. Ivanov, V. Novosad, Y. Otani, H. Shima, and K. Fukamichi, J. Appl. Phys. **91**, 8037 (2002).
[8] K. Y. Guslienko and A. N. Slavin, J. Appl. Phys. **87**, 6337 (2000).
[9] K. Y. Guslienko, A. N. Slavin, V. Tiberkevich, and S. K. Kim, Phys. Rev. Lett. **101**, 247203 (2008).
[10] J. J. Ding, G. N. Kakazei, X. M. Liu, K. Y. Guslienko, and A. O. Adeyeye, Sci. Rep. **4**, 4796 (2014).
[11] F. Boust and N. Vukadinovic, Phys. Rev. B **70**, 172408 (2004).
[12] J. J. Ding, G. N. Kakazei, X. M. Liu, K. Y. Guslienko, and A. O. Adeyeye, Appl. Phys. Lett. **104**, 192405 (2014).
[13] R. Hertel, S. Gliga, M. Fähnle, and C. M. Schneider, Phys. Rev. Lett. **98**, 117201 (2007).
[14] S. O. Demokritov, B. Hillebrands, and A. N. Slavin, Phys. Rep. **348**, 441 (2001).
[15] M. Noske *et al.*, arXiv, http://arxiv.org/abs/1511.03486 (2015).

# Supplementary information for "Three-dimensional Character of the Magnetization Dynamics in Magnetic Vortex Structures - Hybridization of Flexure Gyromodes with Spin Waves"

Matthias Noske,[1,*] Hermann Stoll,[1] Manfred Fähnle,[1] Ajay Gangwar,[2] Georg Woltersdorf,[3] Andrei Slavin,[4] Markus Weigand,[1] Georg Dieterle,[1] Johannes Förster,[1] Christian H. Back[2] and Gisela Schütz[1]

[1]*Max Planck Institute for Intelligent Systems, Heisenbergstr. 3, 70569 Stuttgart, Germany*
[2]*University of Regensburg, Department of Physics, Universitätsstraße 31, 93053 Regensburg, Germany*
[3]*University of Halle, Department of Physics, Von-Danckelmann-Platz 3, 06120 Halle, Germany*
[4]*Oakland University, Department of Physics, Rochester, Michigan 48309, USA*

## Micromagnetic simulations

Our micromagnetic simulations for a circular Permalloy disc with diameter $d = 500$nm and varying thickness $h$ are based on the Landau-Lifshitz-Gilbert equation (LLG) [1], using the object-oriented micromagnetic framework [2] (OOMMF). For the three-dimensional simulations cubic simulations cells with a volume of $(3.125\text{nm})^3$ are used. A finer resolution for the thickness $h$ is achieved by varying the cell size of the simulation in z-direction by $\pm 4\%$. This small variation of the cell size has no detectable influence on the eigenfrequencies as is checked by performing additional simulations for the same thickness but with different cell sizes. Standard material parameters for Permalloy are used for the saturation magnetization $M_S = 690 \text{ kAm}^{-1}$, the exchange constant $A = 13 \cdot 10^{-12} \text{Jm}^{-1}$, gyromagnetic ratio $\gamma = 2.21 \cdot 10^5 \text{mA}^{-1}\text{s}^{-1}$, and anisotropy constants are set to zero. A relatively small damping parameter $\alpha = 0.001$ is used to achieve a high frequency resolution in the Fourier transform analysis (see below), but no notable differences are found in test cases where larger damping constants are used.

The static magnetization $\mathbf{M}_0(\mathbf{r})$ of the vortex in zero-field equilibrium is excited with a short Gaussian in-plane pulse with duration (FWHM) $T = 20$ ps and amplitude $B_0 = 2$ mT, in order to excite the possible eigenmodes. All simulations are performed without an additional external static field. The simulations yield $\mathbf{M}(\mathbf{r}_i, t) = \mathbf{M}_0(\mathbf{r}_i) + \Delta\mathbf{M}(\mathbf{r}_i, t)$ for each discretization point $\mathbf{r}_i$ of the simulation. We consider only the z-component $\Delta M_z(\mathbf{r}_i, t)$ of the dynamic magnetization. To determine the eigenfrequencies $f_N$, a discrete time Fourier transform with a Hanning window and zero padding is performed on $\Delta M_z(\mathbf{r}_i, t)$ for each $\mathbf{r}_i$, yielding $\Delta M_z(\mathbf{r}_i, f)$. From this an amplitude spectrum $I(\mathbf{r}_i, f) = \sqrt{\Delta M_z(\mathbf{r}_i, f)\Delta M_z^*(\mathbf{r}_i, f)}$ is calculated which is averaged over all points $\mathbf{r}_i$, yielding the site-averaged amplitude spectrum $I(f)$ (Fig. I). Each peak of the spectrum at frequency $f_N$ is a candidate for being produced by an eigenmode N with eigenfrequency $f_N$. By resonantly exciting the vortex structure with rotating (CW and CCW) in-plane magnetic fields at the frequencies of these peaks, the eigenmodes can be identified. The frequencies are plotted in Fig. 2 of the paper.

---

[*]Corresponding author: noske@is.mpg.de

The characters of these eigenmodes are analyzed for several thicknesses $h$ of the disc in the following way. We plot perspective representations of the dynamic magnetization $\Delta M_z(\boldsymbol{r_i}, t)$ obtained from the resonant excitation at frequency $f_N$ for an arbitrary $t$ (Fig. 3 of the paper). Furthermore, we produce a snapshot of $\Delta M_z$ for the same time at a cut through the disc at $z = \frac{1}{2}h$ which corresponds to the layer located in the middle of the disc. Finally, we generate diagrams showing the radius $r(z)$ of the gyrotropic motion of the vortex core as function of $z$ (Fig. 2 of the paper). These plots can be interpreted in the following way. If the hybridized eigenmode has contributions from $n = 1, m = \pm 1$ azimuthal spin waves then there is a typical bipolar structure in the planar part of the vortex structure where $\Delta M_z(x_i, y_i)$ is positive in one half of the disc and negative in the other half. This structure is found in all plots of the middle layer of the disc. If the hybridized eigenmode has contributions from $G_n$ flexure modes, then there are $n$ distinct minima in the perspective representation of $\Delta M_z(\boldsymbol{r_i}, t)$ (Fig. 3 b and c of the paper) at the core region and in the plot of the radius $r(z)$ of the gyrotropic motion (Fig. 2b and c of the paper). This does not mean that the gyration radius is exactly zero at the nodes of $G_n$, because there can be contributions from $G_0$.

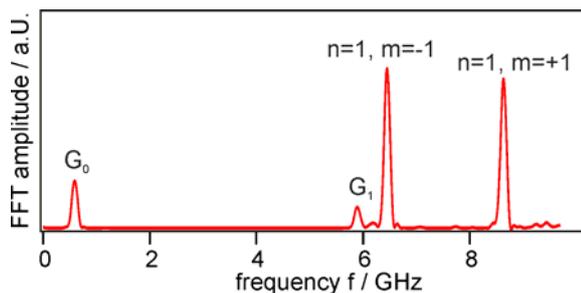

**Figure I**: The site-averaged amplidude spectrum $I(f)$ for a Permalloy disc with diameter $d = 500$ nm and thickness $h = 50$ nm.

# References

[1]    T. L. Gilbert, IEEE Trans. Magn. **40**, 3443 (2004).
[2]    M. J. Donahue and D. G. Porter, *OOMMF User's Guide, Version 1.0, Interagency Report NISTIR 6376, National Institute of Standards and Technology, Gaitherburg, MD.* (1999).